\documentclass[12pt]{article}

\usepackage{graphicx}

\newcommand{\ba}{\begin{eqnarray}}
\newcommand{\ea}{\end{eqnarray}}
\newcommand{\beqs}{\begin{eqnarray}}
\newcommand{\eeqs}{\end{eqnarray}}
\begin{document}
\begin{center}
{\Large\bf \boldmath
Gravimagnetic nucleon form-factors in the impact parameter representation. } 

\vspace*{6mm}
{O.V. Selyugin$^a$ and O.V. Teryaev$^a$ }\\      
{\small \it $^a$ BLTPh, JINR, Dubna \\      
}
\end{center}

\vspace*{6mm}

\begin{abstract}
 In the framework of
 the new t-dependence of the General Parton Distributions (GPDs),
  which reproduce the electromagnetic form factors of
  the proton and neutron at small and large momentum transfer,
the  gravitational form factors of the nucleons and a separate contribution
 of the quarks to them are obtained.
\end{abstract}

\vspace*{6mm}

  As a basis, it is assumed that the form factor is dominated by a soft mechanism
and the Generalized Parton distributions (GPDs)-handbag approach
\cite{R97} is utilized.
 GPDs for $\xi =0 $ provide information
about the distribution of the parton in impact parameter space \cite{Burk00}.
It is connected with $t$-dependence of $GPDs$.


 In \cite{ST1},  a simple ansatz was proposed
which will be good for  describing the form factors of
  the proton and neutron by taking into account a number of new data
that have  appeared in the last years.
We choose
 the $t$-dependence of  GPDs in the form
\ba
{\cal{H}}^{u} (x,t) \  = u(x) \   exp [  a_{+}  \
\frac{(1-x)^2}{x^{m} } \ t ]; \  \  \
{\cal{H}}^{d} (x,t) \  = d(x) \   exp [  a_{+}  \
\frac{(1-x)^2}{x^{m} } \ t ].
\ea
  The size of the parameter $m=0.4$ was determined by the low $t$ experimental data;
 the free parameters $a_{\pm}$ ($a_{+} $ - for ${\cal{H}}$
and $a_{-} $ - for ${\cal{E}}$) were chosen to reproduce the experimental data
in a wide $t$ region.
The  $q(x)$ was taken from
the MRST2002 global fit \cite{MRST02} wth the scale $\mu^2=1 \ $GeV$^2$.
 In all our calculations we restrict ourselves, as in other works, only to
 the contributions of $u$ and $d$ quarks and
 the terms in  ${\cal{H}}^{q}$ and ${\cal{E}}^{q}$.
Correspondingly, for ${\cal{E}}^{u} (x)$,  as for example \cite{R04}, we  have
 \ba
{\cal{E}}^{u} (x) \  = \frac{k_u}{N_u} (1-x)^{\kappa_1} \ u(x),
 \ \ \  \  \
{\cal{E}}^{d} (x) \  = \frac{k_d}{N_d} (1-x)^{\kappa_2} \ d(x),
\ea
 where $\kappa_1 =1.53$ and $\kappa_2=0.31$  \cite{R04}.
With standard normalization of the form factors,
we have
$k_u=1.673, \ \  k_d=-2.033,  \   \ N_u=1.53, \ \  N_d=0.946   $.
The parameters $a_{+} = 1.1$ and $a_{-} $
 were chosen to obtain two possible forms of the ratio of the  Pauli and Dirac form factors.

 \section{Proton and neutron electromagnetic form factors }

 The proton Dirac form factor calculated in \cite{ST1}
  reproduces sufficiently well the behavior of
 experimental data not only at high $t$ but also at  low $t$.
Our description of  the ratio of the Pauli to the Dirac  proton form factors
 and the  ratio of $G_{E}^{p}/G_{M}^{p}$
  shows that in our model we can obtain the results
  of both  the methods (Rosenbluth and Polarization)
by changing the slope of  ${\cal{E} }$.
  Based on the model  developed for proton  the neutron form factors are calculated too.
  To do this the isotopic invariance can be used to change the proton GPDs to neutron GPDs.
 Hence, we do not change any parameters
 and conserve the same $t$-dependence of GPDs as in the case of  proton.
  Our calculation of $G_{E}^{n}$
  shows that the variant which describes the polarization data
   is in better agreement with the  experimental data.
 The calculation of $G_{M}^{n}$
 more clearly  shows that this variant much better describes
 the experimental data especially at low
  momentum transfer.

\section{Gravitational form factors }
   As was shown in \cite{Brodsky00},
the gravitational form  factor for fermions is determined as
\ba
\int^{1}_{-1} \ dx \ x [H(x,\Delta^2,\xi) +E(x,\Delta^2,\xi)] \ \
  = \ \ A_q(\Delta^2)+B_{q}(\Delta^2) .
\ea
\begin{figure}
\begin{flushleft}
\includegraphics[width=.45\textwidth]{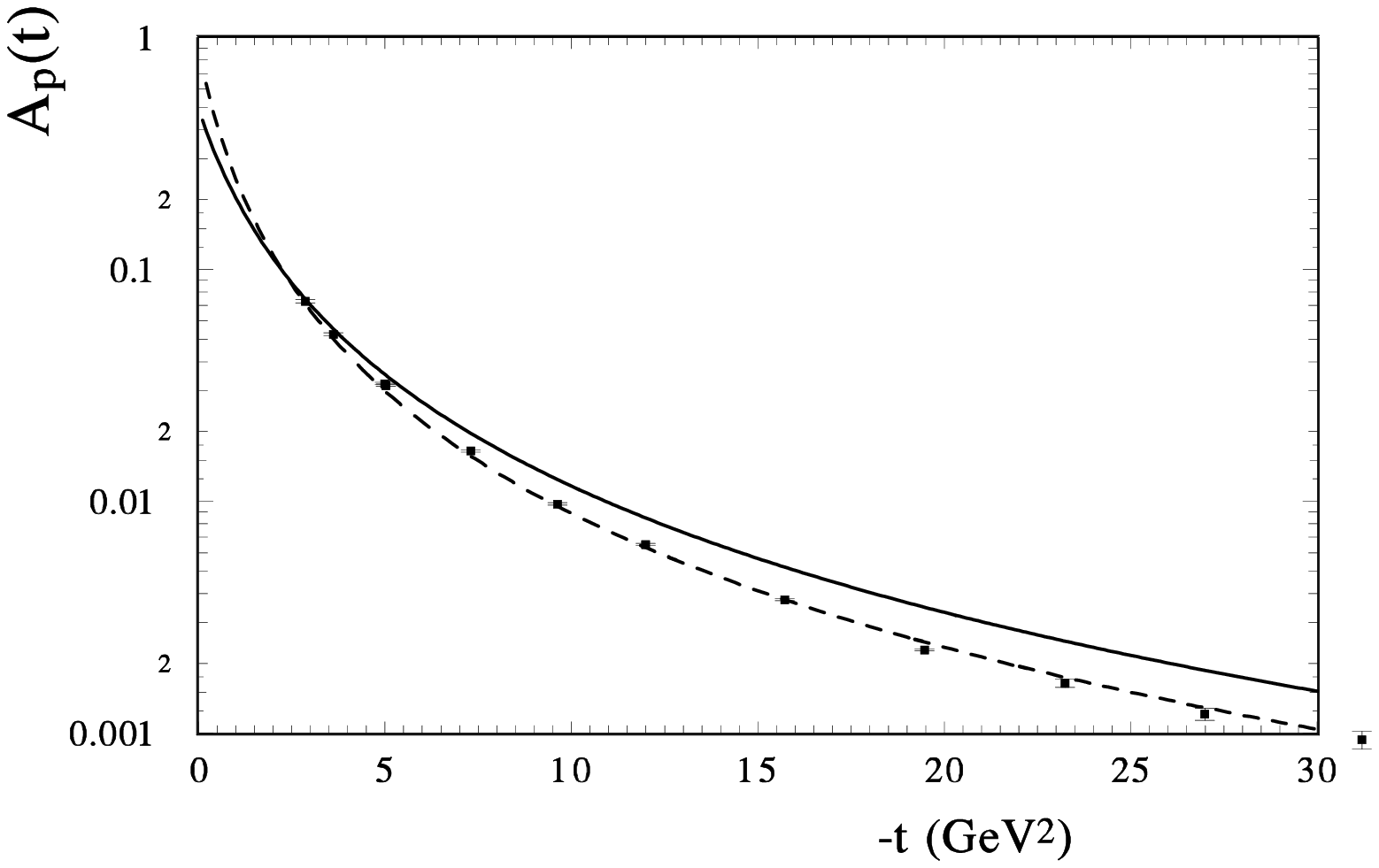}
\end{flushleft}
\vspace{-6.5cm}
\begin{flushright}
\includegraphics[width=.45\textwidth]{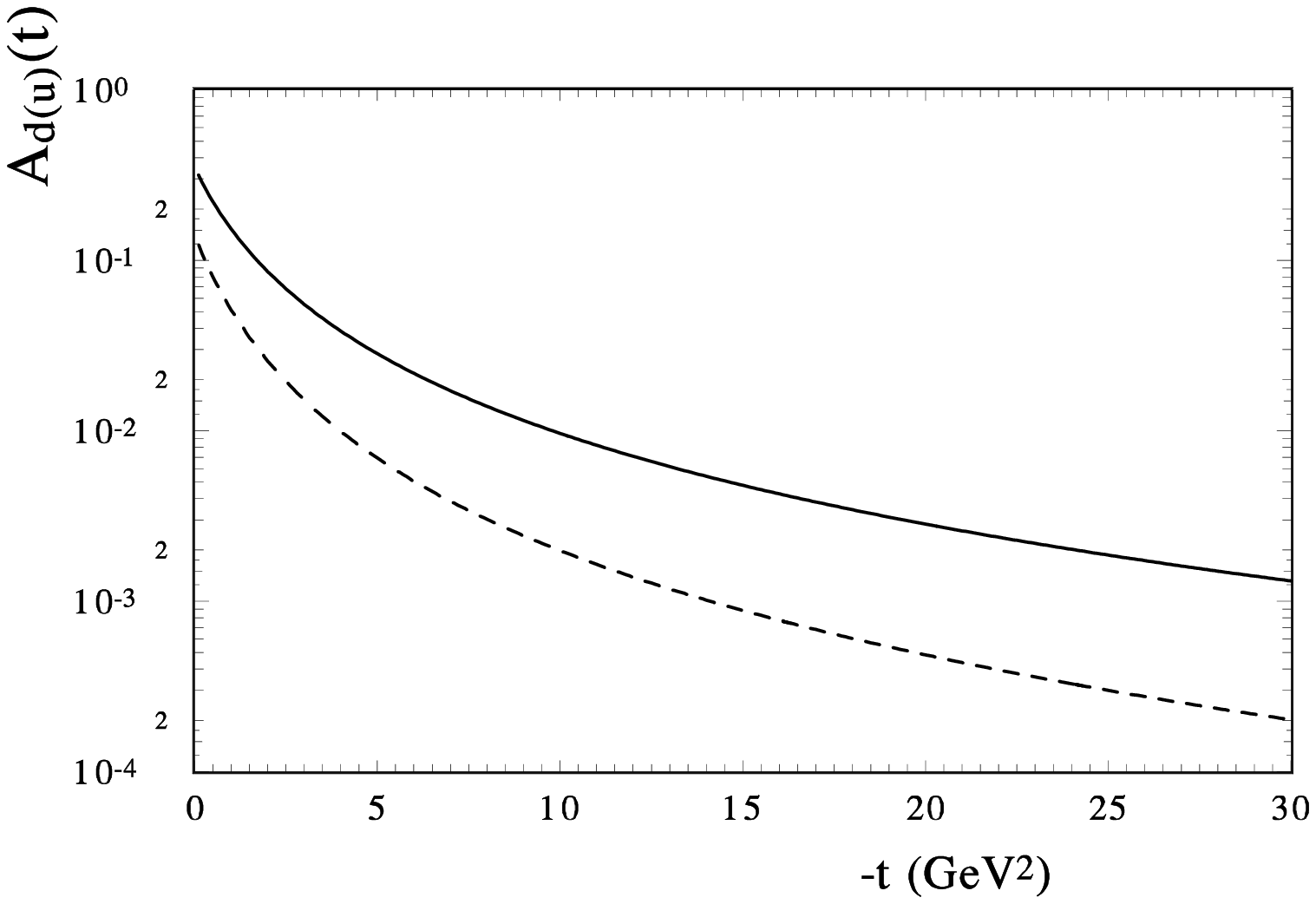}
\end{flushright}
\caption{ a)[left] Gravitation form factor $A_q$ (hard line)
and  Proton Dirac form factor (dashed line),
 both calculated in this work;
 the  data for $F_1^{p}$ are from  \cite{Sill93}.
 b)[right] Contributions of the $u$ (dashed line) and $d$ (hard line)
to the gravitation form factor $A_q$ }
\label{Fig_1}
\end{figure}

 Using this representation we can calculate the gravitational form factor for the nucleon.
 Our result for $A_q(t)$ is shown in Fig.1a. Separate  contributions of the $u$ and $d$-
quark distribution are shown in  Fig.1b.
 At $t=0$ these contributions equal $A_u(t=0)=0.35$
 and $A_d(t=0)=0.14$; and
 $B_u(t=0)=0.22$,
 $B_d(t=0)=-0.27$. The sum of $B_q$ will be near  zero $B_q(t=0)=-0.05$.
In   accuracy of our approximations this result coincides with zero.
 In fig.1a we compare gravitational form factors with our calculations
   of electromagnetic form factors.
 It can be seen that at large momentum transfer they have the same $t$-dependence.
 Of course, they  essentially  differ in  size.

\begin{figure}
\begin{flushleft}
\includegraphics[width=.4\textwidth]{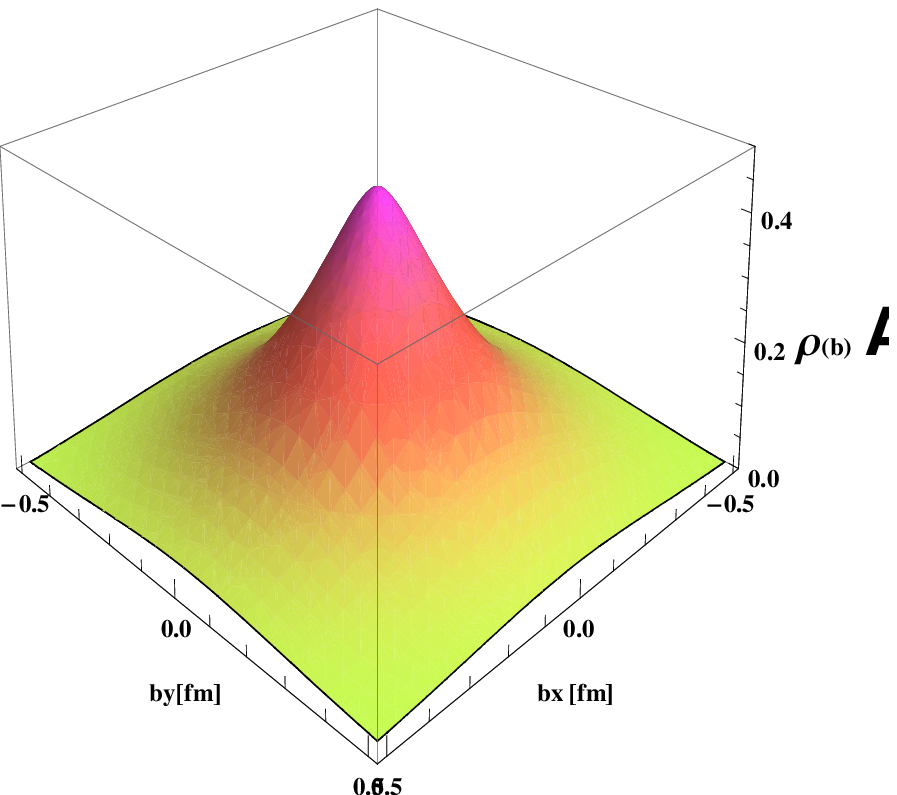} 
\end{flushleft}
\vspace{-6.5cm}
\begin{flushright}
\includegraphics[width=.4\textwidth]{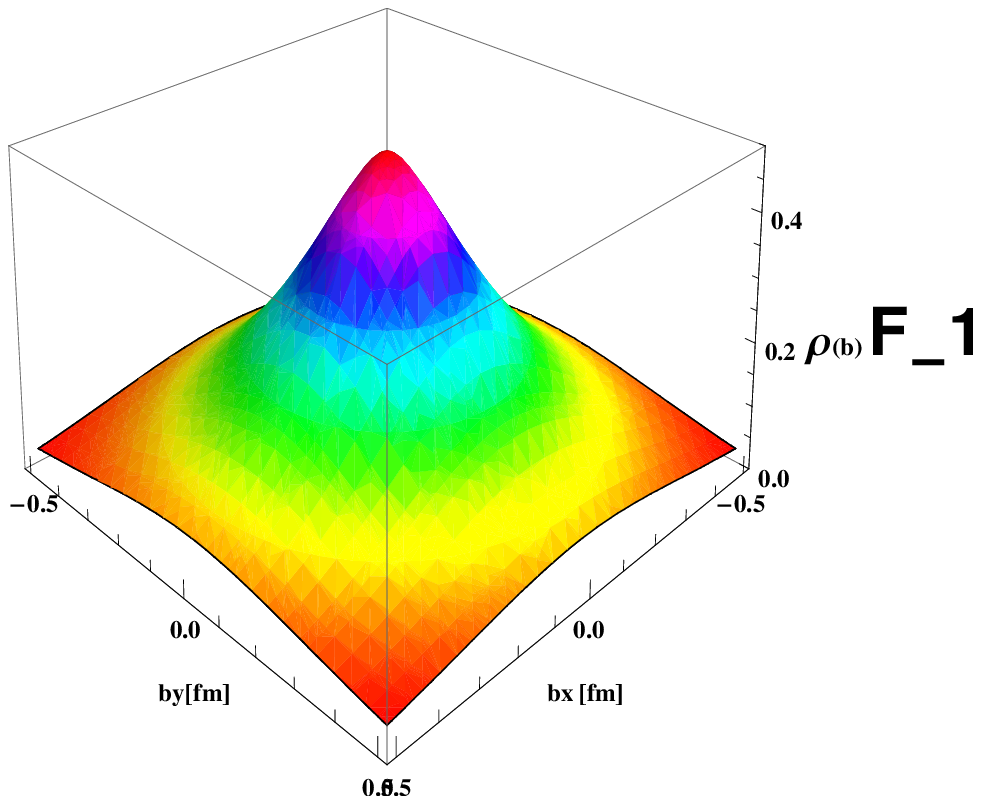} 
\end{flushright}
\caption{Densities of a)[left] the gravitational form factor $A$ and
 b)[right] the electromagnetic  form factor $F_1$.
}
\label{Fig_1}
\end{figure}

\section{Conclusion }
 We introduced a simple new form of the
 $t$-dependence of  GPDs.
 It satisfies the conditions of the non-factorization,
 introduced by Radushkin, and the Burkhardt condition on the power of $(1-x)^n$
 in the exponential form of the $t$-dependence. With this simple form
  we obtained a good description of the proton electromagnetic Sachs form factors.
  Using the isotopic invariance we obtained good descriptions of the neutron
  Sachs form factors without changing any parameters.

   On the basis of our results we calculated the contribution of the $u$ and $d$ quarks
   to the gravitational form factor of the nucleons. The cancellation of these contributions
   at $t=0$ shows that the gravimagnetic form factor is zero for separate contributions,
  gluons and quarks, which confirms the result of \cite{Teryaev-s1}.

\end{document}